\documentclass[10pt,conference]{IEEEtran}

\usepackage{booktabs}
\usepackage{graphicx}
\usepackage[normalem]{ulem}
\usepackage[colorlinks=true, linkcolor=blue, urlcolor=blue, citecolor=blue]{hyperref}
\usepackage{tikz}
\usepackage{amsmath}
\usepackage{svg}
\usepackage{enumitem}
\usepackage{multirow}
\usepackage{mdframed}
\usepackage{cleveref}
\usepackage{algorithm}
\usepackage{algorithmic}
\usepackage{listings}
\usepackage{xcolor}
\usepackage{pifont}
\usepackage{amssymb}
\usepackage{enumitem}

\newlist{breakenum}{enumerate}{1}
\setlist[breakenum]{%
  label=\arabic*.,
  leftmargin=1.5em,
  labelsep=0.5em,
  itemindent=0pt,
  listparindent=\parindent,
  parsep=0pt plus 1pt,
  itemsep=1em plus 0.2em,
  topsep=0pt plus 0.2em,
  partopsep=0pt plus 0.2em,
  align=left,
  before=\leavevmode,
}

\newmdenv[
  linewidth=0.5pt,
  roundcorner=4pt,
  backgroundcolor=gray!10,
  linecolor=gray!50,
  innertopmargin=6pt,
  innerbottommargin=6pt,
  leftmargin=0cm,
  rightmargin=0cm,
  skipabove=10pt,   
  skipbelow=10pt    
]{rqbox}
\definecolor{mycolor}{RGB}{0, 128, 0}
\lstdefinestyle{androidmkstyle}{
    backgroundcolor=\color{white},   
    commentstyle=\color{green},          
    keywordstyle=\color{blue},           
    numberstyle=\tiny\color{gray},       
    stringstyle=\color{red},             
    basicstyle=\ttfamily\scriptsize,     
    breakatwhitespace=false,             
    breaklines=true,                     
    captionpos=b,                        
    keepspaces=true,                     
    numbers=left,                        
    numbersep=5pt,                       
    showspaces=false,                    
    showstringspaces=false,              
    showtabs=false,                      
    tabsize=2,                            
    otherkeywords={:,=},
    morekeywords=[1]{:,=},              
    keywordstyle=[1]\color{blue},       
    morekeywords=[2]{\$},             
    alsoletter={-},
    keywordstyle=[2]\color{red},      
    morekeywords=[3]{include},             
    keywordstyle=[3]\color{mycolor},
    morekeywords=[4]{ture,false,optional},             
    keywordstyle=[4]\color{orange},
    frame=single,        
    rulecolor=\color{black}, 
    framesep=12pt,      
    xleftmargin=10pt, 
    xrightmargin=10pt, 
    aboveskip=1em,     
    belowskip=1em      
}

\AtBeginDocument{%
  }

\usepackage[nolist,nohyperlinks]{acronym}
\acrodef{PII}{Personally Identifiable Information}
\acrodef{GDPR}{General Data Protection Regulation}
\acrodef{CCPA}{California Consumer Privacy Act}
\acrodef{RL}{reinforcement learning}
\acrodef{CP}{Certificate pinning}
\acrodef{TLS}{Transport Layer Security}
\acrodef{CA}{Certificate Authority}
\acrodef{AUT}{Application under test}
\acrodef{PIA}{pre-installed app}
\acrodef{AOSP}{Android Open Source Project}
\acrodef{OEM}{Original Equipment Manufacturer}
\acrodef{ODM}{Original Design Manufacturer}
\acrodef{HAL}{hardware abstraction layer}
\acrodef{OF}{original firmware}
\acrodef{ECF}{emulator compatible firmware}
\acrodef{HAL}{Hardware Abstraction Layer}
\acrodef{GKI}{Generic Kernel Image}
\acrodef{JVM}{Java Virtual Machine}
\acrodef{ART}{Android Runtime}
\acrodef{adb}{Android debug bridge}
\acrodef{AVB}{Android verified boot}
\acrodef{MAC}{Mandatory Access Control}
\acrodef{ABI}{Application Binary Interface}
\acrodef{APK}{Android Package}
\acrodef{APEX}{Android Pony EXpress}
\acrodef{DAC}{Discretionary access control}
\acrodef{QEMU}{Quick Emulator}
\acrodef{AIDL}{Android Interface Definition Language}
\acrodef{UI}{User Interface}
\acrodef{PBI}{Post-Build Injector}
\acrodef{FICD}{Firmware Initialization Completion Detection}
\acrodef{ABI}{Application Binary Interface}

\newcommand\numberOfFirmwareSamples{184}

\newcommand\numberOfLinesModifiedNotExact{300}

\newcommand\numberOfSamplesSDKEins{73}
\newcommand\numberOfSamplesSDKZwei{25}
\newcommand\numberOfSamplesSDKDrei{86}

\newcommand\AvgSuccessRateReHosting{35}

\newcommand\CoveragePercentageSDKEins{31}
\newcommand\CoveragePercentageSDKZwei{19}
\newcommand\CoveragePercentageSDKDrei{42}

\newcommand\MaxCoverageRateReHosting{42}
\newcommand\MinCoverageRateReHosting{19}

\newcommand\AvgBuild{0.97}
\newcommand\AvgBoot{0.97}
\newcommand\AvgSx{68}

\begin{document}
 
\title{Relocate and Emulate: Re-Hosting Android's Application Layer}

\author{\IEEEauthorblockN{1\textsuperscript{st} \href{https://orcid.org/0000-0003-2649-3299}{Thomas Sutter}}
\IEEEauthorblockA{\textit{University of Bern} \\
Bern, Switzerland \\
\href{mailto:thomas.sutter@unibe.ch}{thomas.sutter@unibe.ch}}
\and
\IEEEauthorblockN{2\textsuperscript{nd} \href{https://orcid.org/0000-0002-2582-5557}{Timo Kehrer}}
\IEEEauthorblockA{\textit{University of Bern} \\
Bern, Switzerland \\
\href{mailto:timo.kehrer@unibe.ch}{timo.kehrer@unibe.ch}
\and
\IEEEauthorblockN{3\textsuperscript{rd} \href{https://orcid.org/0000-0002-5008-1107}{Bernhard Tellenbach}}
\IEEEauthorblockA{\textit{Armasuisse Science and Technology} \\
\textit{Cyber-Defense Campus}\\
Thun, Switzerland \\
\href{mailto:bernhard.tellenbach@ar.admin.ch}{bernhard.tellenbach@ar.admin.ch}
}
\and
\IEEEauthorblockN{4\textsuperscript{th} \href{https://orcid.org/0000-0001-5105-3258}{Marc Rennhard}}
\IEEEauthorblockA{\textit{Zurich University of Applied Sciences} \\
Winterthur, Switzerland \\
\href{mailto:marc.rennhard@zhaw.ch}{marc.rennhard@zhaw.ch}
}
}
}

\maketitle
\begin{abstract}
Dynamic analysis of Android’s application layer typically relies on physical devices, limiting scalability and reproducibility. To compensate, we introduce a systematic re-hosting method that relocates the Android framework and pre-installed software from real device firmware into a fully emulated environment. Our approach integrates vendor-specific components into the \ac{AOSP} build system using tailored extraction and injection strategies, producing vendor-flavoured emulator images that preserve system integrity and runtime compatibility. This enables dynamic execution of real-world framework and application-layer components, including proprietary binaries and pre-installed apps, across multiple SDK versions. We evaluate our method on \numberOfFirmwareSamples~firmware samples from SDK 31–33. It achieves high build and boot success rates, with residual failures primarily occurring during core-service initialization due to baseline strategy limitations, missing dependencies, device-protection checks, or emulator constraints. However, the modular design allows injection strategies to be extended for specific firmware, supporting broader compatibility and future research on automated, adaptive re-hosting. Though we identified potential for optimization through engineering vendor-specific solutions, our research demonstrates the feasibility of vendor-flavoured emulators for scalable, reproducible dynamic analysis.
\end{abstract}


\section{Introduction} \label{sec:introduction}
The Android Open Source Project (AOSP) provides a flexible foundation that allows device manufacturers to tailor the operating system to their hardware and market needs. These customizations often involve modifying core components, such as the application framework—including the core system process Zygote~\cite{FromZygotetoMorula}—and integrating proprietary or region-specific applications~\cite{KeepMeUpdated}. While this adaptability enables innovation, it also introduces additional complexity and leads to a fragmented Android ecosystem~\cite{PhDThesisGamba}. Fragmentation poses several risks~\cite{TamingAndroidfragmentation, UnderstandingandDetectingFragmentation}, including the introduction of software bugs~\cite{TheImpactOfVendorCustomizations, KeepMeUpdated}, reduced maintainability~\cite{ELEGANT}, and security vulnerabilities in vendor-specific system components~\cite{KeepMeUpdated, DEFInit, FIRMSCOPE, TowardsUnderstandingAndroid}. 

Despite Android’s open-source foundation, vendor modifications to the application framework make dynamic analysis challenging and often unreproducible~\cite{sutter2024dynamic}. \Acp{PIA} on Android devices illustrate this problem: studies show that they are often outdated, privacy-invasive, or insecure~\cite{FIRMSCOPE, Ananalysisofpreinstalledandroidsoftware, FirmwareDroid, BigMAC}. Most existing analyses, however, rely on static-analysis techniques~\cite{PhDThesisGamba, FirmwareDroid} or are conducted on a limited set of devices~\cite{gamba2020analysis}, which cannot fully capture runtime behaviour. To perform comprehensive dynamic analysis, researchers often need to root or jailbreak devices~\cite{sutter2024dynamic, AndroidRooting}, granting privileged access to system resources and enabling monitoring of otherwise restricted processes. While dynamic analysis provides deeper insight, it is impractical at scale because it requires physical devices that are costly, heterogeneous, and difficult to reproduce due to hardware variability and regional firmware differences~\cite{sutter2024dynamic}.

To compensate, we propose a re-hosting approach that enables dynamic, device-independent analysis of the Android application framework. To our knowledge, this is the first approach that re-hosts the Android application framework from real-world ARM firmware into the Android emulator, enabling vendor-specific dynamic analysis without physical devices. While previous approaches primarily target Linux-based IoT devices~\cite{Pandawan, FIRMADYNE, FirmAE} or RTOS~\cite{FirmPorter}, our work focuses on Android firmware, specifically the re-hosting of application-layer components. Prior systems such as Firmadyne~\cite{FIRMADYNE}, FirmAE~\cite{FirmAE}, and HALucinator~\cite{HALucinator} automate re-hosting for embedded Linux or HAL-level code but do not support Android’s multi-layered framework, binder IPC, or AOSP integration. Our virtualized testing framework supports a wide range of use cases, including security evaluation, software testing, and performance benchmarking.

Our re-hosting method addresses three key challenges. (i) {\em Identifying and extracting vendor-specific framework components}: We analyse firmware from \numberOfFirmwareSamples~smartphones to identify common modification patterns. Our findings show that most vendors extend or modify existing \ac{AOSP} components while largely preserving the original file structure. Based on these insights, we design extraction, and injections routines to reliably retrieve the relevant binaries and resources. (ii) {\em Executing vendor binaries and dependencies in the Android emulator}: We design configurable injection strategies to integrate vendor binaries and their native-library dependencies into an emulator compatible image. Depending on the modification pattern, we apply dynamic injection strategies to ensure correct linking and execution. We demonstrate this by re-hosting the Zygote binary (\texttt{app\_process64}) and other core services, and we measure successful process execution. (iii) {\em Re-hosting custom vendor services}: By integrating into the \ac{AOSP} build process, we create a configurable method for re-hosting vendor-specific services that is applicable across a diverse set of Android devices and OS versions. We leverage \acp{APEX} to dynamically link vendor binaries and their native-library dependencies into an isolated namespace environment. 

We evaluated our re-hosting approach on firmware images from multiple vendors, including Google, Xiaomi, and Qualcomm. For each vendor, we generated functional emulator images based on a unified baseline injection strategy applied across all tested SDK versions. This setup provides a consistent experimental baseline that facilitates reproducible and comparable re-hosting results across vendors. The evaluation confirms that our method is capable to re-host key components of Android’s application layer—including the Zygote process, Java framework, and vendor-specific binaries—while preserving build integrity and system functionality. 

In sum, we present and demonstrate the feasibility of a novel re-hosting method that integrates vendor-specific code into the \ac{AOSP} build process, producing vendor-flavoured emulator images compatible with the Android emulator. These images enable scalable, dynamic testing of application-layer components in a fully emulated environment.
While we adopted a baseline configuration that is intentionally conservative for reproducibility in terms of our experiments, vendor-tailored optimizations create a clear path towards further improving re-hosting coverage and runtime success in future work. Through systematic evaluation of file coverage ratios and success rates, we establish a foundation for targeted enhancements and adaptive re-hosting strategies.  

We summarize our main contributions as follows:

\begin{enumerate}[left=0.25em]
    \item A systematic, build-integrated method for re-hosting Android’s application layer from vendor firmware into the emulator.

    \item Automated extraction, signing, and a base set of configurable injection techniques for integrating application-layer components.

    \item Empirical validation showing functional vendor-flavoured emulator images across SDK 31–33 with a generic baseline injection strategy.

    \item An open, reproducible infrastructure enabling scalable dynamic analysis of real-world Android firmware with study artefacts publicly available at: \\\href{https://sites.google.com/view/relocate-and-emulate}{https://sites.google.com/view/relocate-and-emulate}
\end{enumerate}

\section{Background} \label{sec:Background}

\begin{figure*}[t]
    \centering
    \includegraphics[width=\linewidth]{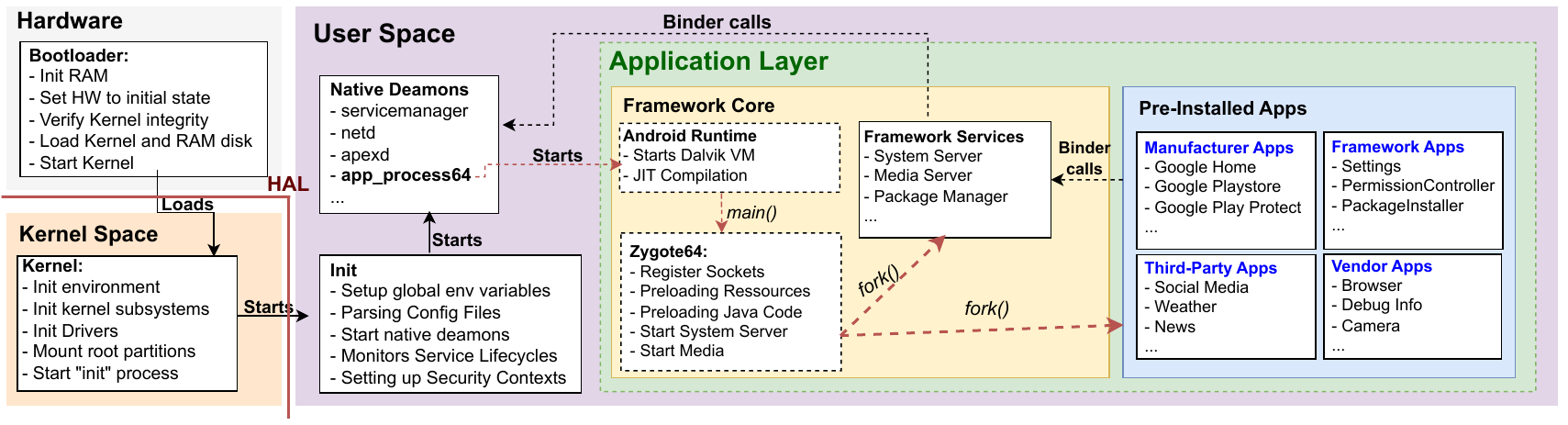}
    \vspace{-5mm}
    \caption{Android architecture with logical layers.}
    \label{fig:01_Implementation-Layer}
\end{figure*}

The Android software architecture is organized into several distinct layers, as illustrated in Figure~\ref{fig:01_Implementation-Layer}. At its foundation lies the Linux kernel, which provides essential system functionality such as process and memory management, device driver support, and security enforcement. Above the kernel resides the Hardware Abstraction Layer (HAL), a collection of standardized interfaces that enables the Android framework to communicate with hardware components—such as GPS, radio, and sensors—without requiring knowledge of their device-specific implementations. Built upon this foundation is the user space, which comprises the Android application layer. This layer includes the software components responsible for managing and executing user-facing applications and system services, forming the visible interface between the Android system and the end user.

While this layered design promotes portability and modularity, it also creates dependency chains between components that can complicate re-hosting efforts. Vendor-specific modifications can occur across all layers—from kernel patches and custom HAL modules to modified framework services and proprietary applications—introducing additional complexity when attempting to reproduce system behaviour in a re-hosted environment. In our work, we focus solely on the application layer, as it encapsulates the Android framework and pre-installed applications.

Focussing on the application layer, Android uses multiple disk partitions (e.g., \textit{system}, \textit{vendor}, \textit{system\_ext}, and \textit{product}) to organize the operating system and its components. These partitions serve to separate privileges and responsibilities between system and vendor code, allowing device manufacturers to customize the OS without compromising compatibility with the Android framework. In practice, the application layer spans over several disk partitions and in theory, remains isolated from the kernel and hardware via the HAL.

The application layer consists primarily of framework bytecode files (\texttt{.dex}, \texttt{.jar}), system services (\texttt{.elf} , \texttt{.apex}), native libraries (\texttt{.so}), static resources (\texttt{.xml}, \texttt{.json}, \texttt{.ogg}, etc.), and pre-installed apps (\texttt{.apk}, \texttt{.odex}, \texttt{.vdex}). While this list is not exhaustive and may vary across products and device classes (e.g,. smartphones, cars, etc.), our re-hosting approach focuses on extracting and integrating these key components into a working emulator image to enable dynamic analysis. To do so, we implemented custom routines to handle Android specific file formats, such as the \ac{APEX} file format.

\section{Challenges}\label{sec:Challenges}
Android applications are primarily developed in Java or Kotlin and compiled into Dalvik Executable (DEX) bytecode—a CPU architecture-independent format used by the Android Runtime (ART). In addition to DEX code, many apps include native C/C++ libraries that must be compiled for the target architecture (typically ARM). Most Android devices today use ARM-based processors, making native components incompatible with x86/x86\_64-based emulators commonly used in previous research~\cite{Bringingbalancetotheforce, sinha2020emulation, AppJitsu}. To address this, we developed our approach specifically for ARMv8a and use an ARM-compatible emulator to ensure compatibility with native code.

Dynamic analysis of \acp{PIA} is constrained by Android’s security architecture, which enforces system integrity and app isolation across multiple layers:
\begin{itemize}[left=0.25em]
\item \textbf{SELinux Policies:} Mandatory access control (MAC) rules restrict process privileges. Customizations by the \acp{OEM} often diverge from AOSP defaults, causing vendor apps to fail under stricter policies.

\item \textbf{Read-Only Partitions:} Privileged apps must reside on partitions such as \texttt{/system} or \texttt{/product}, which are protected by \ac{AVB} and cannot be modified without breaking the verified boot chain.

\item \textbf{App Signing:} System apps require valid platform or media certificates; signature mismatches lead to installation rejection or secure-boot failures.

\item \textbf{Permission Whitelisting:} Privileged permissions are statically defined in XML configuration files on read-only partitions, preventing dynamic modification at runtime.
\end{itemize}

Beyond these protections, several technical challenges further complicate re-hosting of application layer components:
\begin{itemize}[left=0.25em]
\item \textbf{Singleton Apps:} Certain core services (e.g., the permission controller) exist only once in the system; duplicates cause conflicts or boot failures.

\item \textbf{Inter-App Dependencies (Collusion):} Many \acp{PIA} rely on shared system services or Binder IPC; missing dependencies lead to crashes.

\item \textbf{File and Library Dependencies:} Vendor apps depend on firmware-specific native libraries and configuration files that must be correctly restored.

\item \textbf{APEX Dual Signing:} \ac{APEX} modules encapsulate signed mini-filesystems verified with both vendor and platform keys. Re-hosting requires extracting, validating, and re-signing both layers to maintain integrity.
\end{itemize}

Taken together, these challenges demonstrate that simply transferring a core binary, or \acp{PIA} to a generic emulator is insufficient for meaningful re-hosting. A careful reconstruction of its runtime context—including security policy alignment, filesystem layout, and dependency resolution—is essential.

\section{Approach}\label{sec:Methodology}
\begin{figure}[t]
    \centering
    \includegraphics[width=\linewidth]{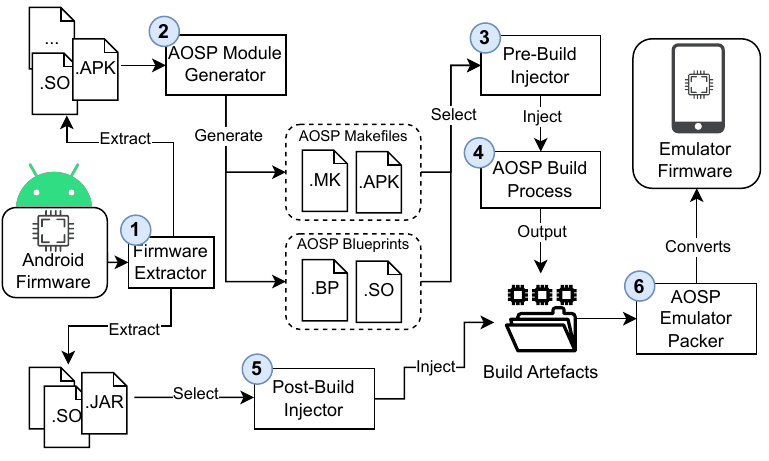}
    \vspace{-5mm}
    \caption{Overview of the re-hosting workflow, leveraging the AOSP build process and vendor firmware to create a vendor-flavoured emulator image.}
    \label{fig:MotivationAOSPBuild}
\end{figure}

Our goal is to re-host Android's framework to enable the dynamic analysis of core services and \acp{PIA} created by Android device vendors (e.g., Google, Xiaomi, etc.). Figure~\ref{fig:MotivationAOSPBuild} illustrates the core idea of our re-hosting approach. Leveraging Android’s build process, we create an emulator image that includes key components of the vendor firmware—framework files, native libraries, services, and PIAs—while preserving system integrity and compatibility with the Android emulator. To achieve this, we propose a multi-stage process comprising six steps.

\paragraph*{\textbf{Step 1: Firmware Extraction}}
In this step, the \textit{Firmware Extractor} unpacks vendor-provided Android firmware images and systematically identifies all components belonging to the application-layer, including framework files, native libraries, binaries, static assets, and \acp{PIA}.

\paragraph*{\textbf{Step 2: AOSP Module Generation}}
The \textit{AOSP Module Generator} takes the extracted files as input and creates AOSP-compatible build descriptors. Android supports two formats for build module definitions: "\textit{Android.mk}" and "\textit{Android.bp}". The generator dynamically creates these files using metadata extracted from the firmware (e.g., file location, signing key, permissions).
Listing~\ref{lst:androidmk} shows an example of an "\textit{Android.mk}" file that integrates a prebuilt APK. Key build variables such as {\footnotesize \texttt{LOCAL\-\_MODULE\-\_PATH}}, {\footnotesize \texttt{LOCAL\-\_CERTIFICATE}}, and {\footnotesize \texttt{LOCAL\-\_PRIVILEGED\-\_MODULE}} must be set correctly to ensure proper signing, placement, and privilege assignment. Otherwise, integrity checks would stop the boot process of the device.


\paragraph*{\textbf{Step 3: Pre-Build Injection}}
The generated modules are placed into the AOSP source tree without disrupting the build process. This involves placing the modules in the appropriate directories, updating build configuration files to include the new modules, and ensuring that no build rule conflicts with existing emulator targets.

\paragraph*{\textbf{Step 4: AOSP Build Process Wrapper}}
Invoking the AOSP build system produces a system image that includes the extracted vendor packages, ensuring that all extracted components are correctly signed (e.g., platform, network stack, media keys), prebuilt components are preserved without unnecessary recompilation, files are mounted in the expected filesystem paths (e.g., \texttt{/system\-/priv-app/}), and dependencies to native libraries are correctly resolved. 

\paragraph*{\textbf{Step 5: Post-Build Injection}}
Not all components can be injected via the standard build system. For instance, pre-compiled `.jar` or binary `.elf` to replace core components are not supported by the build system. Thus, such files are injected after the base AOSP build into the file system. The post-processing step adds or replaces existing files within the image, ensures that \ac{AVB} integrity is preserved, and updates file permissions and ownerships to match runtime expectations.

\paragraph*{\textbf{Step 6: AOSP Emulator Packer}}
Finally, we produce a vendor-flavoured emulator image, manually configuring the build to disable unnecessary emulator-specific features (e.g., telephony mocks or stub modules), ensure a minimal bootable system with the required runtime services, and validate that re-hosted applications and system services function correctly at runtime.



\section{Implementation} \label{sec:Implementation}
In this section, we describe how we realize our Android re-hosting approach within a research prototype for the sake of experimentation.
In Section~\ref{sec:impl-steps}, we detail the extraction and integration of vendor-specific components into the AOSP build system to produce functional emulator images. Our implementation prioritizes automation, reproducibility, and minimal AOSP modifications, while using flexible injection strategies to handle diverse vendor firmware. In Section~\ref{sec:impl-cross}, we address cross-cutting key considerations such as signing, dependency resolution, and build stability to ensure correct execution of re-hosted components.

\subsection{Implementation of the Major Steps}
\label{sec:impl-steps}

\paragraph*{\textbf{Step 1: Firmware Extraction}}
Our system supports the extraction of application-layer files from key partitions found in vendor firmware images: \texttt{super}, \texttt{system}, \texttt{system\_ext}, \texttt{vendor}, and \texttt{product}. For this task, we leverage \textit{FirmwareDroid}~\cite{FirmwareDroid}, a firmware analysis framework that supports unpacking major Android versions across a wide range of vendors and extend its extraction capabilities for our purposes. We extract all relevant files and retrieve static metadata including signing certificates, file paths, permissions, and user/group identifiers. This metadata is essential for the accurate generation of AOSP-compatible build modules.

\paragraph*{\textbf{Step 2: AOSP Module Generator}}
Since the emulator’s filesystem layout does not always align with that of the target firmware, we map vendor file paths to the closest semantically equivalent location in the emulator image (e.g., \texttt{/system/system/\-app} to \texttt{/system\-/app}). This ensures structural compatibility and preserves logical organization. Additionally, based on the \texttt{android:\-sharedUser\-Id} value declared in each app’s \texttt{Android\-Manifest.xml}, we select and reference the appropriate AOSP signing key (e.g., \texttt{platform}, \texttt{media}) in the generated module configuration.

The output of the build module generator is a collection of prebuilt AOSP modules either Makefiles (\texttt{.mk}) or Blueprint files (\texttt{.bp})—one for each app or native library. All remaining file types (e.g., binaries, static assets, or firmware blobs) are handled during the post-build injection phase.

\paragraph*{\textbf{Step 3: Pre-Build Injection}}
The \textit{Pre-Build Injector} is responsible for integrating the generated build modules into the AOSP source tree. It places modules into the appropriate partition source directory and links them into the build dynamically by adjusting core build files (e.g., base\_system.mk). Each module is assigned to a target partition (e.g., \texttt{system}, \texttt{vendor}) based on its original location in the vendor firmware.

A key limitation of this approach lies in the inability to overwrite certain core AOSP components with pre-built modules. For instance, attempting to inject a prebuilt \texttt{framework\--res.apk} will typically result in a duplicate module conflict and trigger build errors. Patching the AOSP source code to resolve such conflicts is technically possible, but would compromise the portability and generality of our method.

To handle such cases without modifying AOSP source files, we implement a set of fallback strategies:
\begin{itemize}[left=0.25em]
    \item \textbf{Overwrite Directive:} Use the AOSP \texttt{LOCAL\-\_OVERRIDE\-\_BUILT\-\_MODULES} directive to prioritize our prebuilt version.
    
    \item \textbf{Post-Build Injection:} Defer integration to the post-build injection phase, bypassing the AOSP build system entirely.
    
    \item \textbf{Module Removal or Patching:} Remove the conflicting AOSP module or patch its definition to avoid duplication only in cases where the module is not system relevant.
\end{itemize}

\begin{figure}[t]
\lstset{style=androidmkstyle} 
\begin{lstlisting}[
  caption={Android make file example.},
  label={lst:androidmk},
]
LOCAL_PATH := $(call my-dir)
include $(CLEAR_VARS)
LOCAL_MODULE_TAGS := optional 
LOCAL_MODULE := Example
LOCAL_MODULE_PATH := $(TARGET_OUT)/system/app/
LOCAL_CERTIFICATE := platform
LOCAL_SRC_FILES := Example.apk
LOCAL_MODULE_CLASS := APPS
LOCAL_MODULE_SUFFIX := $(COMMON_ANDROID_PACKAGE_SUFFIX)
LOCAL_OPTIONAL_USES_LIBRARIES := 
LOCAL_ENFORCE_USES_LIBRARIES := false
LOCAL_DEX_PREOPT := false
LOCAL_PRIVILEGED_MODULE := false

include $(BUILD_PREBUILT)
\end{lstlisting}
\end{figure}

\paragraph*{\textbf{Step 4: AOSP Build Process Wrapper}}
To support our method, certain modifications to the AOSP source code are necessary. Overall, the total amount of AOSP code changes required is less than \numberOfLinesModifiedNotExact~lines. These modifications primarily disable specific AOSP features or adjust build configurations to make the build process more stable and efficient. Re-hosting framework binaries, such as Zygote's \texttt{app\_process64}, presents significant challenges due to their direct dependencies on numerous native libraries and the enforcement of security checks, as discussed in Section~\ref{sec:Background}. Native libraries often have transitive dependencies on other libraries, which further complicates integration.

Consequently, when re-hosting a core binary, it is essential to include both its direct and indirect dependencies to ensure all necessary symbols are available for correct execution. To address this, we implement configurable injection strategies that allow for custom rules tailored to different binaries and firmware.

\paragraph*{\textbf{Step 5: Post-Build Injection}}
After the AOSP build completes, each partition (e.g., \texttt{system}, \texttt{vendor}) is represented by a dedicated directory containing its corresponding output artifacts. This structure allows us to inspect, modify, and extend the build output before the image-packing phase. In this stage, we selectively add pre-compiled \texttt{.jar} or native \texttt{.so} files extracted from the vendor firmware, or replace existing ones where appropriate.

To determine which artifacts should be injected or overwritten, we iterate through all files extracted from the vendor firmware and apply the matching Algorithm~\ref{alg:FileMatchingAlgorithm}. The algorithm compares each vendor file against intermediate AOSP build artifacts using attributes such as Application Binary Interface (ABI), file type, source and target location, file extension, module name, and privilege level. This matching process is essential, as many intermediate build files have generic names (e.g., \texttt{package.apk}, \texttt{javalib.jar}) and are dispersed across multiple directories in the AOSP build tree. By identifying the correct intermediate targets, we ensure that vendor components are accurately integrated into the final emulator image without disrupting the build process.

\begin{algorithm}[htb]
\caption{Matching vendor firmware files to AOSP intermediate files}
\label{alg:FileMatchingAlgorithm}
\footnotesize
\begin{algorithmic}[1]
\REQUIRE List of vendor files $V$, List of AOSP intermediate files $A$
\STATE Initialize empty list of matches $M$
\FOR{each $v \in V$}
    \STATE Initialize best match to None, best match $a_{\text{best}} \leftarrow \texttt{None}$
    \FOR{each $a \in A$}
        \STATE Compute match based on ABI, file type, path similarity, module name, privileges, file name, module type, keywords
        \IF{match}
            \STATE Update match $a_{\text{best}} \leftarrow a$
        \ENDIF
    \ENDFOR
    \IF{$a_{\text{best}} \neq \texttt{None}$}
        \STATE Add $(v, a_{\text{best}})$ to match list $M$
    \ENDIF
\ENDFOR
\RETURN $M$
\end{algorithmic}
\end{algorithm}

The result of Algorithm~\ref{alg:FileMatchingAlgorithm} is a list with files in the vendor firmware image that have a counterpart in the AOSP intermediate files. With this information, we can then decide on the injection strategy for each file in the vendor firmware image. 

\paragraph*{\textbf{Injection Strategies}}
We define four types of injection strategies depending on the characteristics and compatibility of the vendor file:

\begin{itemize}[left=0.25em]
    \item \textbf{Indirect Injection:} If a corresponding file already exists in the build output (e.g., an APEX or JAR file), we apply a matching algorithm to locate its intermediate source and overwrite it with the vendor version.
    
    \item \textbf{Direct Injection:} If no equivalent file exists in the output directories, we copy the vendor file into the corresponding partition directory. We replicate the original path from the vendor firmware to preserve directory structure and compatibility. 

    \item \textbf{Isolated Namespace Injection:} Binaries that must run in an isolated namespace are encapsulated within an APEX containing all necessary dependencies. The binary is replaced with a symbolic link pointing to the APEX version, ensuring that the dynamic linker resolves dependencies only from the packaged libraries. Inclusion of binaries is performed manually by the user via an explicit list.

    \item \textbf{No Injection:} Files that cannot be injected safely (e.g., incompatible kernel modules) are skipped and logged. These are typically files that conflict with the base kernel or violate signing constraints.
\end{itemize}

All injection strategies are configurable and customizable to support a wide variety of vendors and devices. However, depending on the file to inject dependency and security constraints have to be considered. 

\paragraph*{\textbf{Step 6: AOSP Emulator Packer}}
The final build step in AOSP is the image-packing process. During this stage, intermediate files are copied into the partition directories and overwritten by any differing output artifact. By modifying intermediate files and placing additional files using our injection routines, we ensure that the resulting emulator image includes all vendor-specific components required for testing. Our modifications to AOSP are minimal (less than \numberOfLinesModifiedNotExact~lines of code) and do not break the integrity of the AOSP source, making the method portable across different Android versions.

\subsection{Cross-Cutting Considerations}
\label{sec:impl-cross}

\paragraph*{\textbf{Pipeline Reproducibility and Automation}}
For the emulator images, we used the \texttt{sdk\_phone64\_arm64\--userdebug} AOSP build target. This configuration produces a minimal ARM64-based image that includes the Android SDK, a design choice aimed at maximizing compatibility across a wide range of Android smartphone samples. Supporting other device classes (e.g., Android Automotive) would require selecting an appropriate build target for the respective platform.

\paragraph*{\textbf{Android Integrity and Signature Mechanisms}}
APK files are signed using the certificates defined in the AOSP build configuration. APEX containers require dual (or, in some cases, trial) signing: one signature for the internal payload file system and another for the APEX container itself. Additionally, if the APEX package includes an embedded APK, that APK must also be individually signed~\cite{APEXfileformat}. If any of these signatures are invalid or missing, the Android system will either reject the module or fail to boot entirely. Thus, correct signing is essential for multiple steps of the pipeline.

Our signing routines automate this multi-layered signing process by leveraging AOSP's existing key infrastructure. In particular, these routines allow us to overwrite the contents of APEX files without breaking their integrity. This is essential as APEX packages often contain core system components required for the OS to start, such as the Android Runtime. This design allows for any modifications that do not violate Android’s integrity mechanisms (e.g., signature mismatches, critical path conflicts).

\section{Evaluation Setup}\label{sec:Evaluation_Setup}
This section describes the setup used to evaluate the feasibility and effectiveness of our re-hosting approach for Android firmware. We focus on systematically assessing how well application-layer components can be re-hosted and executed in the emulator. Our evaluation is guided by the following research questions:
\begin{description}[]   
    \item[\textbf{RQ1}] To what extent can application-layer components — including framework files, native libraries, and pre-installed apps — be automatically re-hosted?

    \item[\textbf{RQ2}] Can Android’s vendor-specific application-layer components be successfully re-hosted and executed within an emulator environment?
    
    \item[\textbf{RQ3}] What is the computational and build-time overhead introduced by the re-hosting process compared to a baseline AOSP build?
    
    \item[\textbf{RQ4}] Under what conditions does the proposed re-hosting approach fail, and what are the root causes of these failures?
\end{description}
Overall, \textbf{RQ1} assesses how much of the Android application layer can be re-hosted automatically, indicating the achievable analysis coverage and scalability to large firmware sets. \textbf{RQ2} evaluates whether vendor-specific components execute correctly in an emulator, which is essential for reliable dynamic analysis without physical devices. \textbf{RQ3} measures the computational and build-time overhead of re-hosting, informing its practicality for large-scale and continuous analysis pipelines. Finally, \textbf{RQ4} analyses failure modes and their root causes, clarifying the limitations of the approach and guiding future improvements.

To address these questions, we assess (i) the compatibility of the method across diverse firmware images, (ii) the coverage of application-layer components that can be successfully injected and executed, and (iii) the runtime behaviour of re-hosted applications in the emulator environment. 


\subsection{Experimental Subjects}
We apply our prototype to a dataset of real-world firmware images from multiple vendors (e.g., Google, Xiaomi, etc.), collected from websites that freely provide official firmware downloads; the complete dataset is included in our replication package.
In sum, we extract application-layer files from \numberOfFirmwareSamples~samples, which serve as the basis for our evaluation. All samples in our dataset are based on the SDK versions 31 to 33. The tests were conducted using the official Android emulator for ARM64 (v33.1.24.0). Runtime experiments were performed on an ARM-based host system, while firmware builds were executed on Intel x86\_64 based machines equipped with 64\,GB of RAM and 64 vCPU cores. 
Formally, we denote the dataset by
 \[
 \mathcal{D} = \{ x_1, x_2, \dots, x_{n} \},
 \]
where $n$ is the total number of samples. 
The dataset is partitioned into three disjoint subsets according to the SDK version:
\begin{align*}
\mathcal{D}_{31} &= \{ x \in D \mid \text{SDK}(x) = 31 \}, & \mathcal{E}_{31} &= \text{android-12.0.0\_r34},\\
\mathcal{D}_{32} &= \{ x \in D \mid \text{SDK}(x) = 32 \}, & \mathcal{E}_{32} &= \text{android-12.1.0\_r11},\\
\mathcal{D}_{33} &= \{ x \in D \mid \text{SDK}(x) = 33 \}, & \mathcal{E}_{33} &= \text{android-13.0.0\_r16}.
\end{align*}

Here, each subset $\mathcal{D}_i$ corresponds to samples built from the Android build target $\mathcal{E}_i$. The build targets were selected by Android SDK version. The build target defines the exact AOSP source version used to construct the system image for the corresponding API level. 
To demonstrate the capabilities of our approach, we re-host the \texttt{Zygote} process and core framework components (i.e., \path{/system/framework}, \path{/vendor/framework}, \path{/product/framework}, and \path{/system_ext/framework}). We define the set of binaries to be re-hosted as
\[
\mathcal{B} = \{b_1, b_2, \dots, b_k \},
\]
where each \(b_i\) represents a re-hosted binary component of the system. We denote the set of partitions containing framework components by
\[
\mathcal{P} = \{\text{/system}, \text{/vendor}, \text{/product}, \text{/system\_ext}\}.
\]
The set of framework components can then be defined as
\[
\mathcal{F} = \{ F \mid F \text{ is a framework directory in some partition } P \in \mathcal{P} \}.
\]
In addition to the framework, our approach re-hosts a selection of native libraries and pre-installed Android applications. We denote these sets by
\[
\mathcal{L} = \{\ell_1, \ell_2, \dots, \ell_m\}, \quad
\mathcal{A} = \{a_1, a_2, \dots, a_n\},
\]
where $\ell_i$ represents a native library and $a_j$ a pre-installed application. The complete set of re-hosted components is therefore
\[
\mathcal{C} = \mathcal{B} \cup \mathcal{F} \cup \mathcal{L} \cup \mathcal{A}.
\]

\subsection{Injection Strategy}
\label{sec:Evaluation_InjectionStrategy}
All the conducted experiments use a single, well-defined baseline injection strategy applied to each sample \(x \in \mathcal{D}\). Injection is performed according to these rules: (i) \(\mathcal{B}\) (binaries/APEX) are injected via the post build injector; APEX files are either directly injected as vendor APEXes or merged with emulator APEX files and for every image we inject the Android Runtime and SDK-Extension APEX files; (ii) \(\mathcal{F}\) (framework) components overwrite existing emulator components. \(\mathcal{A}\) (Android applications) are injected directly only if they do not already exist on the target partition, otherwise they are applied indirectly (i.e., via build-time integration or overlay); (iii) \(\mathcal{L}\) native libraries are injected only if absent on the target system. 

We deliberately restricted our evaluation to this single strategy in order to establish a reproducible, automation-friendly baseline for re-hosting. Exploring alternative injection heuristics with the goal of maximizing per-sample success and coverage rates is beyond the scope of this work. Nevertheless, the strategy can be adapted on a per-device basis where necessary to support a wider range of vendor firmwares, and we view the systematic exploration of such optimizations as promising future work.

\subsection{Evaluation Metrics}
Our evaluation metrics for assessing the re-hosting capabilities in our experiments are inspired by related work on firmware re-hosting (see Section~\ref{sec:RelatedWork}), adapted to our context of the Android application layer. 

For {\bfseries RQ1}, we measure, for each sample $x \in \mathcal{D}$, the coverage of application-layer components that are successfully re-hosted.  
Let $\mathcal{C}_x \subseteq \mathcal{C}$ denote the subset of components from $\mathcal{C}$ that are successfully re-hosted on sample $x$.  
Further, let $\mathcal{I}_x \subseteq \mathcal{C}$ denote the set of all components from the same sample that could in principle be injected. Note that the number of files that could be injected depends on how many files were successfully extracted from a sample. The per-sample coverage ratio is then defined as
\[
R(x) = \frac{|\mathcal{C}_x|}{|\mathcal{I}_x|}.
\]

Similarly, for a subset of samples $\mathcal{D}_i \subseteq \mathcal{D}$, the average coverage ratio is
\[
R(\mathcal{D}_i) = \frac{1}{|\mathcal{D}_i|} \sum_{x \in \mathcal{D}_i} R(x).
\]

To evaluate the success of executing re-hosted components within an emulator environment ({\bfseries RQ2}), we define a set of tasks to be executed. The first task is to generate an AOSP compatible firmware that can be {\em built} in AOSP and loaded by the emulator. The second task measures whether the system successfully {\em boots} the OS. We define a successful boot as moment when the Android Debug Bridge (adb) is started and available. The third task assesses whether the re-hosted framework {\em core} services initialize correctly without crashing. In particular, if Zygote and the Android Runtime start correctly without fatal crashes. For the fourth tasks, we define success in case the user-interface is shown via a {\em launcher} app and usable in the emulator. Let the set of tasks to evaluate the re-hosted system be
\[
\mathcal{T} = \left\{
\begin{array}{l}
\text{build success}, \text{boot success}, \\ 
\text{init core services}, \text{init launcher app}
\end{array}
\right\}.
\]
For each sample \(x \in \mathcal{D}\), we define a success function
\[
s(x, t) = 
\begin{cases}
1, & \text{if task } t \in \mathcal{T} \text{ succeeds on sample } x,\\
0, & \text{otherwise}.
\end{cases}
\]

The overall success rate of a sample can then be expressed as
\[
S(x) = \frac{\sum_{t \in \mathcal{T}} s(x, t)}{|\mathcal{T}|}.
\]

Similarly, the average success rate across a subset of samples \(\mathcal{D}_i \subseteq \mathcal{D}\) is
\[
S(\mathcal{D}_i) = \frac{1}{|\mathcal{D}_i|} \sum_{x \in \mathcal{D}_i} S(x).
\]

Finally, for {\bfseries RQ3}, we compare the running times in our build pipeline to unmodified baseline AOSP builds, and resort to a qualitative failure analysis for {\bfseries RQ4}.

\section{Experimental Results}\label{sec:Results}

This section presents the empirical evaluation of our re-hosting approach using the firmware dataset introduced in Section~\ref{sec:Evaluation_Setup}. The goal is to assess its feasibility, correctness, and efficiency across different SDK versions and vendors. We report results along four dimensions corresponding to our research questions: coverage of application-layer components successfully integrated into the emulator, correctness and runtime behaviour of re-hosted services and pre-installed apps, computational overhead introduced by the injection process, and root causes of remaining failures.

\begin{figure}
    \centering
    \includegraphics[width=\linewidth]{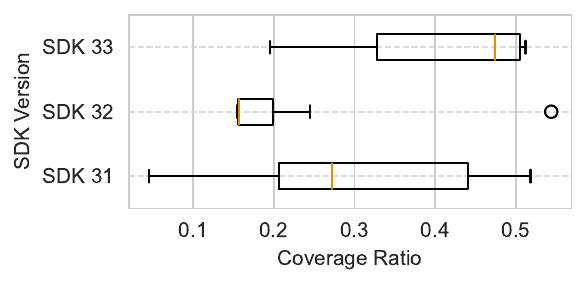}
    \vspace{-5mm}
    \caption{Application-Layer Coverage Ratios}
    \label{fig:coverage}
\end{figure}

\subsection{RQ1: Coverage of Application-Layer Files}
\label{sec:coverage}

Using the baseline injection strategy, our experiments achieved an average coverage, \( R(\mathcal{D}_i) \),  of \CoveragePercentageSDKEins\% for \(\mathcal{D}_{31}\) (sample size 73), \CoveragePercentageSDKZwei\% for \(\mathcal{D}_{32}\) (sample size 25), and \CoveragePercentageSDKDrei\% for \(\mathcal{D}_{33}\) (sample size 86). Figure~\ref{fig:coverage} shows the distribution of application-layer coverage across all samples. As shown, the baseline injection strategy attains the highest coverage for SDK~33, while the lowest coverage was achieved for SDK-32.

The observed variation in coverage is influenced by differences in vendor modifications, framework complexity, and the presence of device-specific dependencies across SDK versions. SDK~33 shows higher coverage, likely due to more consistent file structures and fewer missing dependencies in the analysed samples. Conversely, SDK~31 presents less heterogeneous layouts, which challenge the baseline strategy. These findings highlight the trade-off between a general, reproducible injection approach and the potential benefits of device-specific tuning.

\begin{rqbox}
\textbf{RQ1:} The average coverage across SDK versions ranges from \MinCoverageRateReHosting\% to \MaxCoverageRateReHosting\%, with the highest value observed for SDK~33.
\end{rqbox}

\subsection{RQ2: Correctness of Re-hosting}
\label{sec:correctness}



We measure the ability of the re-hosting system to perform the set of evaluation tasks \(\mathcal{T}\) on the dataset \(\mathcal{D}\). The re-hosting of the components \(\mathcal{C}\) is applied individually to each subset 
\(\mathcal{D}_{31}\), \(\mathcal{D}_{32}\), and \(\mathcal{D}_{33}\). The results for the average success rate are shown in Table~\ref{tab:firmware-samples}. 

Using our re-hosting approach, we successfully re-hosted core components across multiple devices and vendors. The consistently high build and boot success rates under the baseline strategy demonstrate that our combination of pre- and post-build injection is robust and preserves the integrity of the AOSP build system. For example, despite challenges such as complex signing requirements and singleton app constraints described in Section~\ref{sec:Challenges}, our method consistently maintained system integrity across all tested SDK versions. These results show that the challenges discussed in Section~\ref{sec:Challenges} can be effectively addressed by our approach.

We observed that the launcher task generally followed similar success trends as the core service initialization. Notably, while the launcher application typically started, it did not always function as expected. In some cases, the user interface exhibited transient flickering, likely caused by repeated failures of other services that disrupted execution. These observations suggest that, although the underlying system services and the launcher were successfully initialized, subtle timing or rendering inconsistencies within the emulator environment can affect launcher stability, revealing opportunities for refinement in post-build injection and runtime configuration.

\begin{rqbox}
\textbf{RQ2:} Our results demonstrate that vendor-specific components can be reliably re-hosted in an ARM emulator, achieving correct core initialization for \AvgSuccessRateReHosting\% of the samples and an average success rate of \AvgSx\%.
\end{rqbox}

\begin{table}[t]
\centering
\footnotesize
\caption{Correctness of re-hosting: Task success within the emulator.}
\label{tab:firmware-samples}
\renewcommand{\arraystretch}{1.5}
\begin{tabular}{|c|c|c|c|c|c|c|}
\hline
\multicolumn{2}{|c|}{} & \multicolumn{5}{c|}{\textbf{Task Success}} \\
\hline
\textbf{Subset} & \textbf{\#Sample} & 
\textbf{Build} & \textbf{Boot} & \textbf{Core} & \textbf{Launch} & 
\textbf{Avg.\ $S(\mathcal{D}_i)$}  \\
\hline
$\mathcal{D}_{33}$ & \numberOfSamplesSDKDrei & 1.0 & 1.0 & 0.41 & 0.41 & 0.70 \\
\hline
$\mathcal{D}_{32}$ & \numberOfSamplesSDKZwei & 1.0 & 1.0 & 0.68 & 0.68 & 0.80 \\
\hline
$\mathcal{D}_{31}$ & \numberOfSamplesSDKEins & 0.9 & 0.9 & 0.18 & 0.18 & 0.54 \\
\hline
\textbf{Overall} & \textbf{\numberOfFirmwareSamples} & \textbf{\AvgBuild} & \textbf{\AvgBoot} & \textbf{0.\AvgSuccessRateReHosting} & \textbf{0.\AvgSuccessRateReHosting} & \textbf{0.\AvgSx} \\
\hline
\end{tabular}
\end{table}

\subsection{RQ3: Performance Overhead}\label{sec:performance}
\begin{figure}[htbp]
    \centering
    \includegraphics[width=\linewidth]{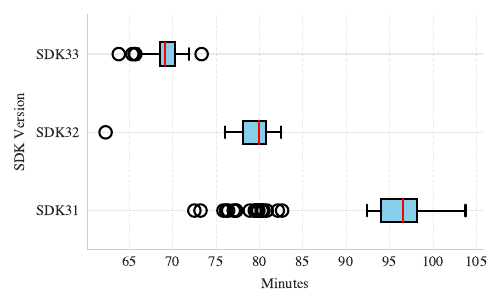}
    \vspace{-5mm}
    \caption{AOSP Build Times with Injected Modules}
    \vspace{-5mm}
    \label{fig:aosp_build_times}
\end{figure}

\begin{figure}[htbp]
    \centering
    \includegraphics[width=\linewidth]{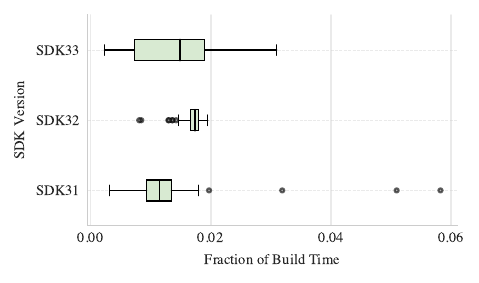}
    \vspace{-5mm}
    \caption{Post Injector Build Times in Fraction of the Build Time}
    \label{fig:post_injection_build_times}
\end{figure}

We measure the time required for the injection steps in our build pipeline and compare it against a baseline AOSP build for each API version without any modifications. Figure~\ref{fig:aosp_build_times} illustrates the distribution of build times for each SDK version across all samples. A default AOSP build without modifications takes approximately 60~minutes for SDK~33, 75~minutes for SDK~32, and 90~minutes for SDK~31 to complete on our test systems. As expected, build times increase when additional AOSP modules are injected. On average, one build with vendor modules injected requires only slightly more time than the baseline. The module pre-injection phase completes in less than three minutes and newer SDK versions have generally faster build times. Furthermore, Figure~\ref{fig:post_injection_build_times} shows that the post-injection step accounts for only a small fraction of the total build time, as Algorithm~\ref{alg:FileMatchingAlgorithm} operates with linear time complexity. Overall, the injection process introduces only minimal computational overhead, completing on average in under three minutes.

\begin{samepage}
\begin{rqbox}
\textbf{RQ3:} The pre- and post-injection stages add less than three minutes of overhead, confirming that injection performance costs are negligible for large-scale use.
\end{rqbox}
\end{samepage}

\begin{table*}[htb]
\centering
\scriptsize
\noindent
\caption{Summary of platform, app, and security capabilities and limitations.\\ \checkmark = full support, (\checkmark) = partial support, \ding{55} = no support.}
\vspace{-3mm}
\label{tab:contributions}
\begin{minipage}[t]{0.48\linewidth}
\vspace{0pt} 
\centering
\begin{tabular}{@{}ll@{}}
\toprule
\multicolumn{2}{l}{\textbf{Platform Capabilities / Limitations}} \\ \midrule
Singleton App Support \checkmark & Supports one active app instance. \\
System File Installation \checkmark & Correct installation into system paths (e.g., /priv-app). \\
Core Framework Re-hosting \checkmark & Full re-hosting of Java framework components. \\
File \& Native-Library Support (\checkmark) & Selective re-hosting of system files and native libraries. \\
Daemon Re-hosting (\checkmark) & Partial re-hosting of essential native daemons. \\ \midrule
Kernel Module Support \ding{55} & Kernel modules are not re-hosted. \\
Hardware Emulation (\checkmark) & Default emulator hardware device emulation. \\ \midrule

\multicolumn{2}{l}{\textbf{App-Level Capabilities / Limitations}} \\ \midrule
Inter-App Dependence \checkmark & Supports colluding or dependent apps. \\
App File \& Library Support (\checkmark) & Selective re-hosting of app files and libraries. \\ \midrule
External Service Dependence \ding{55} & No support for external API dependencies. \\
Anti-Analysis Mitigation \ding{55} & Does not prevent app anti-analysis or env-checks. \\
\bottomrule
\end{tabular}
\end{minipage}
\hfill
\begin{minipage}[t]{0.48\linewidth}
\vspace{0pt} 
\centering
\begin{tabular}{@{}ll@{}}
\toprule
\multicolumn{2}{l}{\textbf{Security Capabilities / Limitations}} \\ \midrule
Verified Boot (AVB) \checkmark & Passes all AVB security checks. \\
Read-Only File Systems \checkmark & Maintains read-only partitions at runtime. \\
App Integrity Protection \checkmark & All apps signed with correct platform keys. \\
APEX Integrity Protection \checkmark & APEX files signed and verified. \\
Permission Whitelisting (\checkmark) & Re-hosts permissions and logs violations. \\
Device Root Access \checkmark & Full root and unrestricted kernel control. \\ \midrule
SELinux Enforcement (\checkmark) & Only permissive mode supported. \\
Device Attestation \ding{55} & Attestation fails (due to permissive SELinux). \\
\bottomrule
\end{tabular}
\end{minipage}
\end{table*}

\subsection{RQ4: Failure Analysis}
\label{sec:failure_analysis}
Our approach achieved consistently high build and boot success rates across all SDK subsets (see Table~\ref{tab:firmware-samples}). Nonetheless, a number of failures persisted, primarily during core-service initialization. To better understand these residual issues, we conducted a root-cause analysis and summarised the results in Table~\ref{tab:contributions}. The analysis revealed that failures mainly stem from (i) limitations of the baseline injection strategy, (ii) missing or unresolved native dependencies, (iii) integrity verification errors triggered by device-protection mechanisms, and (iv) limited emulator support for ARM~32-bit routines or peripherals (e.g., key-master interface). The first three categories represent engineering limitations that can be mitigated by refining the injection and signing procedures or by improving dependency resolution, whereas the fourth reflects an inherent constraint of the emulator platform rather than a deficiency of our approach. Consequently, the observed failures delineate practical—rather than conceptual—boundaries of re-hosting, underscoring the robustness and generality of the proposed method.

\begin{samepage}
\begin{rqbox}
\textbf{RQ4:} Re-hosting failures primarily occur during core service initialization due to missing symbols, integrity checks, or limited emulator support.
\end{rqbox}
\end{samepage}

\section{Discussion} \label{sec:Discussion}
In this section, we summarize the key observations from our evaluation wrt.\ to our primary evaluation goal of demonstrating the general feasibility of our approach (Section~\ref{sec:feasibility}), before we discuss the major threats to validity of our experimental results (Section~\ref{sec:threats}).

\subsection{Feasibility of the Approach}
\label{sec:feasibility}

\paragraph*{\textbf{Relocation and emulation}} 
Pre- and post-injection steps around our AOSP build process wrapper represent a crucial and configurable aspect of our re-hosting method. The generic baseline injection strategy chosen for the experiments was deliberately conservative to ensure reproducibility. Despite this, a significant portion of key components is re-hosted, and device-specific tuning could further increase coverage ratios and thus execution success rates within the emulator.
    
Core framework services and pre-installed apps successfully boot and initialize, demonstrating that vendor binaries and framework files can be re-hosted while preserving system integrity, and thus vendor-specific Android application-layer components can be dynamically executed in an emulator. This unlocks the potential for vendor-flavoured emulators, empowering scalable and reproducible dynamic analysis, which is currently not possible within the Android ecosystem.

Failures arise from unresolved native dependencies, integrity verification errors, or emulator constraints (e.g., missing peripherals), representing mainly practical, not conceptual, limits.

\paragraph*{\textbf{Performance overhead}} For the conducted experiments, pre- and post-build injection completes in under three minutes, and full vendor-flavoured builds average between 60 to 95 minutes. 
Compared to baseline AOSP builds without any modifications, this represents a moderate overhead of around 5\% on average, without observing any significant outliers. While both performance optimization and systematic performance benchmarking are out of the scope this paper, our performance measurements further demonstrate the applicability of our approach in practice.

\paragraph*{\textbf{Maintainability and portability}}
Our method is designed to be largely generic and not tightly coupled to any specific vendor firmware layout. By relying on the standard AOSP build system and focusing solely on application-layer components, we maintain broad compatibility and portability across different devices and SDK versions.
Our research prototype shows that our approach may be implemented with minimal modifications to the AOSP source. Nonetheless, adjustments to injection strategies may be required to support newer Android versions.

\paragraph*{\textbf{Findings}}
Our results show that vendor-specific Android application-layer components can be re-hosted and dynamically executed in an emulator. Across \numberOfFirmwareSamples~firmware images from multiple vendors and SDK versions, we achieve high build and boot success rates and demonstrate correct initialization of core framework services under a single, reproducible baseline strategy. These findings establish emulator-based re-hosting as a practical and scalable foundation for dynamic analysis, while also revealing dependency resolution and integrity checks as the primary remaining challenges.

\subsection{Threats to Validity}
\label{sec:threats}

We discuss threats to validity following standard guidelines~\cite{wohlin2012experimentation}, distinguishing between internal, construct, external, and conclusion validity.

\paragraph{Internal Validity}
Our evaluation uses a single, uniform baseline injection strategy to ensure reproducibility.
While this limits confounding factors, it may reduce success and coverage rates, as device-specific tuning could resolve additional dependencies or configuration issues.
Residual failures during core service initialization may also stem from extraction or signing artefacts.
Furthermore, emulator-based execution cannot fully capture hardware-dependent behaviour (e.g., sensors, peripherals, or ARM 32-bit components), which may affect runtime correctness.

\paragraph{Construct Validity}
We measure re-hosting success using build and boot success, initialization of core services and the launcher, and coverage ratios of re-hosted application-layer components.
These metrics provide a practical approximation of functional correctness but do not guarantee full behavioural equivalence to physical devices, for example with respect to timing or subtle service interactions.

\paragraph{External Validity}
Our dataset comprises 184 firmware images from SDK versions 31--33 and may not represent all Android devices, vendors, or OS versions.
Nevertheless, it includes multiple major vendors and recent Android releases, capturing realistic variation in vendor-specific framework and application-layer modifications.

\paragraph{Conclusion Validity}
All samples are evaluated using a consistent experimental setup, reducing random effects and selective bias.
Although absolute success rates depend on the conservative baseline strategy and emulator limitations, the observed trends across vendors and SDK versions consistently support our conclusions on feasibility, scalability, and reproducibility.


\begin{table*}[t]
\centering
\footnotesize
\caption{Comparison of firmware re-hosting approaches.}
\renewcommand{\arraystretch}{1}
\begin{tabular}{|l|c|c|c|c|c|c|c|}
\hline
\textbf{Feature / Metric} & \textbf{Pandawan} & \textbf{Firmadyne} & \textbf{FirmAE} & \textbf{HALucinator} & \textbf{Our Work} \\
\hline
Target OS / Platform & Linux & Linux & Linux & Android & Android \\
Focus & IoT & IoT & IoT & HAL & Framework \\
Vendor-specific support & No & No & No & Yes & Yes \\
Hardware Modelling Required & No & Yes & Yes & Yes & No \\
Scalability / Automation & High & High & High & Medium & High \\
Full System Re-Hosting & Yes & Yes & Yes & No & No \\
\hline
\end{tabular}
\label{tab:comparison}
\end{table*}

\section{Related Work} \label{sec:RelatedWork}

Research on firmware analysis and system emulation has produced many frameworks for reverse engineering, vulnerability discovery, and behavioural analysis. Our work intersects with but differs from prior approaches in firmware re-hosting~\cite{FIRMADYNE, FirmAE, FirmDiff, SURGEON}. Our approach addresses a unique problem space: the re-hosting of vendor-modified Android application frameworks within the AOSP build system to generate emulator-compatible images. This design enables reproducible and scalable dynamic analysis of framework-level behaviour without relying on physical devices, manual hardware modelling, or peripheral pass-through. 

Several works address automated re-hosting of Linux-based firmware to facilitate dynamic analysis. Firmadyne~\cite{FIRMADYNE} pioneered large-scale emulation of embedded Linux firmware, focusing primarily on network appliances such as routers and cameras. 
It leverages QEMU for full-system emulation and automates filesystem extraction, but does not support Android-specific frameworks or services. FirmAE~\cite{FirmAE} extends this concept for large-scale IoT firmware emulation, 
improving automation and scalability.
FirmDiff~\cite{FirmDiff} focuses on refining Linux kernel configurations to improve compatibility during firmware re-hosting,
reducing emulation failures due to module layout mismatches. The work of Wright et al.~\cite{ChallengesInFirmwareReHosting} surveys the broader challenges in firmware re-hosting, highlighting hardware modelling, dependency resolution, and scalability as recurring obstacles. Pandawan~\cite{Pandawan} proposes a benchmarking methodology to quantify progress in Linux-based firmware re-hosting, 
introducing the \ac{FICD} metric. Pretender \cite{Pretender} automates re-hosting by dynamically inferring hardware interfaces and adapting the emulation environment accordingly.
SURGEON~\cite{SURGEON} advances re-hosting by introducing a binary-rewriting approach to re-host functional components from one firmware image into another.
These efforts significantly advance the state of firmware re-hosting but largely target general-purpose embedded Linux environments. Android frameworks introduce additional challenges, such as service initialization ordering, binder IPC dependencies, and AOSP build integration, which are not addressed by these systems.

Binary analysis frameworks like Angr~\cite{angr} provide deep static and dynamic program analysis at the application or binary level, while HALucinator~\cite{HALucinator} facilitates selective re-hosting of system components by modelling HAL code; however, neither approach targets the integration of vendor-specific Android frameworks nor compatibility with the AOSP build system, which is the primary focus of our work.

Table~\ref{tab:comparison} summarises the comparison of our re-hosting approach with the most closely related firmware re-hosting systems, including Pandawan~\cite{Pandawan}, Firmadyne~\cite{FIRMADYNE}, FirmAE~\cite{FirmAE}, and HALucinator~\cite{HALucinator}.  Unlike Firmadyne, FirmAE, and HALucinator, our approach does not require hardware modelling, which improves scalability and simplifies automation. Vendor-specific support is provided, similar to HALucinator, enabling evaluation across multiple device manufacturers. Our evaluation focuses on success per component or task, which provides a quantitative measure of functionality, while HALucinator focuses on coverage, i.e., how much of the system is represented. Finally, although full system re-hosting is supported by most existing approaches, our work emphasizes selective component re-hosting, particularly the framework and application-layer components, to facilitate modular analysis and testing of this components.

\section{Conclusion} \label{sec:Conclusion}
We presented a systematic approach for re-hosting Android system components and applications in an emulator, maintaining compatibility across core services, framework binaries, and pre-installed apps while preserving device integrity. Our evaluation demonstrates that critical services and vendor-specific binaries can be executed successfully, resulting in a fully operational emulator that mirrors the behaviour of original hardware and supports scalable dynamic analysis.

Overall, our re-hosting strategy provides a reproducible and extensible framework for exploring Android system internals, performing security analyses, and investigating vendor-specific modifications in a controlled emulator environment. While this work establishes the foundations for dynamic execution of vendor components, future work will focus on adapting specific dynamic analysis workflows—such as automated UI interaction, runtime monitoring of framework and pre-installed app behaviour, and system call or network tracing. These targeted analyses will enable reproducible end-to-end studies of vendor-specific services and pre-installed apps in a controlled and scalable setting. In addition, future work will focus on extending compatibility across diverse vendor images, integrating advanced injection strategies, and further improving correctness and performance fidelity.
In summary, our results show that emulator-based re-hosting enables reliable testing of pre-installed apps, evaluation of vendor frameworks, and large-scale dynamic analysis, while laying the foundation for broader coverage, enhanced compatibility, and automated evaluation of Android firmware.

 \appendix
 \section*{A: Data Availability}\label{sec:Availability}
Artefacts available:
\begin{itemize}
    \item \url{https://sites.google.com/view/relocate-and-emulate}
\end{itemize}

\section*{Acknowledgements}
This work was supported in part by armasuisse Science and Technology.

\bibliographystyle{IEEEtran}
\bibliography{biblio}

@article{ChallengesInFirmwareReHosting,
author = {Wright, Christopher and Moeglein, William A. and Bagchi, Saurabh and Kulkarni, Milind and Clements, Abraham A.},
title = {Challenges in Firmware Re-Hosting, Emulation, and Analysis},
year = {2021},
issue_date = {January 2022},
publisher = {Association for Computing Machinery},
address = {New York, NY, USA},
volume = {54},
number = {1},
issn = {0360-0300},
url = {https://doi.org/10.1145/3423167},
doi = {10.1145/3423167},
journal = {ACM Comput. Surv.},
month = {jan},
articleno = {5},
numpages = {36}
}

@inproceedings{HALucinator,
  author    = {Abraham A. Clements and Eric Gustafson and Tobias Scharnowski and Paul Grosen and David Fritz and Christopher Kruegel and Giovanni Vigna and Saurabh Bagchi and Mathias Payer},
  title     = {{HALucinator}: Firmware Re-hosting Through Abstraction Layer Emulation},
  booktitle = {Proceedings of the 29th USENIX Security Symposium (USENIX Security '20)},
  year      = {2020},
  isbn      = {978-1-939133-17-5},
  pages     = {1201--1218},
  publisher = {USENIX Association},
  address   = {Boston, MA, USA},
  month     = aug,
  url       = {https://www.usenix.org/conference/usenixsecurity20/presentation/clements}
}

@inproceedings{FIRMSCOPE,
  author    = {Mohamed Elsabagh and Ryan Johnson and Angelos Stavrou and Chaoshun Zuo and Qingchuan Zhao and Zhiqiang Lin},
  title     = {{FIRMSCOPE}: Automatic Uncovering of Privilege-Escalation Vulnerabilities in Pre-Installed Apps in Android Firmware},
  booktitle = {Proceedings of the 29th USENIX Security Symposium (USENIX Security '20)},
  year      = {2020},
  isbn      = {978-1-939133-17-5},
  pages     = {2379--2396},
  publisher = {USENIX Association},
  address   = {Boston, MA, USA},
  month     = aug,
  url       = {https://www.usenix.org/conference/usenixsecurity20/presentation/elsabagh}
}

@inproceedings{FirmwareDroid,
  author    = {Thomas Sutter and Bernhard Tellenbach},
  title     = {FirmwareDroid: Towards Automated Static Analysis of Pre-Installed Android Apps},
  booktitle = {Proceedings of the 2023 IEEE/ACM 10th International Conference on Mobile Software Engineering and Systems (MOBILESoft)},
  year      = {2023},
  pages     = {12--22},
  publisher = {IEEE/ACM},
  address   = {Melbourne, Australia},
  doi       = {10.1109/MOBILSoft59058.2023.00009}
}

@inproceedings{SURGEON,
  author    = {Florian Hofhammer and Marcel Busch and Qinying Wang and Manuel Egele and Mathias Payer},
  title     = {SURGEON: Performant, Flexible, and Accurate Re-Hosting via Transplantation},
  booktitle = {Proceedings of the Workshop on Binary Analysis Research (BAR'24)},
  year      = {2024},
  publisher = {Internet Society},
  address   = {Reston, VA, USA},
  month     = mar,
  doi       = {10.14722/bar.2024.23011},
  url       = {https://www.ndss-symposium.org/ndss-paper/auto-draft-430/},
  pages     = {1--10}, 
  numpages  = {10} 
}

@inproceedings{Ananalysisofpreinstalledandroidsoftware,
  author    = {Julien Gamba and Mohammed Rashed and Abbas Razaghpanah and Juan Tapiador and Narseo Vallina-Rodriguez},
  title     = {An Analysis of Pre-Installed Android Software},
  booktitle = {Proceedings of the 2020 IEEE Symposium on Security and Privacy (SP)},
  year      = {2020},
  address   = {San Francisco, CA, USA},
  pages     = {1039--1055},
  publisher = {IEEE}
}

@inproceedings{Pretender,
author = {Eric Gustafson and Marius Muench and Chad Spensky and Nilo Redini and Aravind Machiry and Yanick Fratantonio and Davide Balzarotti and Aur{\'e}lien Francillon and Yung Ryn Choe and Christophe Kruegel and Giovanni Vigna},
title = {Toward the Analysis of Embedded Firmware through Automated Re-hosting},
booktitle = {22nd International Symposium on Research in Attacks, Intrusions and Defenses (RAID 2019)},
year = {2019},
isbn = {978-1-939133-07-6},
address = {Chaoyang District, Beijing},
pages = {135--150},
url = {https://www.usenix.org/conference/raid2019/presentation/gustafson},
publisher = {USENIX Association},
month = sep
}

@inproceedings {Pandawan,
author = {Ioannis Angelakopoulos and Gianluca Stringhini and Manuel Egele},
title = {Pandawan: Quantifying Progress in Linux-based Firmware Rehosting},
booktitle = {33rd USENIX Security Symposium (USENIX Security 24)},
year = {2024},
isbn = {978-1-939133-44-1},
address = {Philadelphia, PA},
pages = {5859--5876},
url = {https://www.usenix.org/conference/usenixsecurity24/presentation/angelakopoulos},
publisher = {USENIX Association},
month = aug
}

@inproceedings{FirmDiff,
  author    = {Ioannis Angelakopoulos and Gianluca Stringhini and Manuel Egele},
  title     = {FirmDiff: Improving the Configuration of Linux Kernels Geared Towards Firmware Re-hosting},
  booktitle = {Proceedings of the Workshop on Binary Analysis Research (BAR'24)},
  year      = {2024},
  editor    = {NDSS Program Committee},
  publisher = {Internet Society},
  address   = {San Diego, CA, USA},
  month     = mar,
  doi       = {10.14722/bar.2024.23012},
  url       = {https://www.ndss-symposium.org/wp-content/uploads/bar2024-12-paper.pdf},
  pages     = {1--10},  
  numpages  = {10}     
}

@inproceedings{FirmAE,
author = {Kim, Mingeun and Kim, Dongkwan and Kim, Eunsoo and Kim, Suryeon and Jang, Yeongjin and Kim, Yongdae},
title = {FirmAE: Towards Large-Scale Emulation of IoT Firmware for Dynamic Analysis},
year = {2020},
isbn = {9781450388580},
publisher = {Association for Computing Machinery},
address = {New York, NY, USA},
url = {https://doi.org/10.1145/3427228.3427294},
doi = {10.1145/3427228.3427294},
booktitle = {Proceedings of the 36th Annual Computer Security Applications Conference},
pages = {733–745},
numpages = {13},
location = {Austin, USA},
series = {ACSAC '20}
}

@inproceedings{FIRMADYNE,
  author    = {Daming D. Chen and Maverick Woo and David Brumley and Manuel Egele},
  title     = {Towards Automated Dynamic Analysis for Linux-based Embedded Firmware},
  booktitle = {Proceedings of the Network and Distributed System Security Symposium (NDSS)},
  year      = {2016},
  pages     = {1--10},  
  numpages  = {10},
  publisher = {Internet Society},
  address   = {San Diego, CA, USA},
  month     = feb,
  doi       = {10.14722/ndss.2016.23004},
  url       = {https://www.ndss-symposium.org/wp-content/uploads/2017/09/towards-automated-dynamic-analysis-linux-based-embedded-firmware.pdf}
}

@inproceedings{angr,
  author    = {Fish Wang and Yan Shoshitaishvili},
  title     = {Angr: The Next Generation of Binary Analysis},
  booktitle = {Proceedings of the 2017 IEEE Cybersecurity Development Conference (SecDev)},
  year      = {2017},
  pages     = {8--9},
  publisher = {IEEE},
  address   = {San Francisco, CA, USA},
  doi       = {10.1109/SecDev.2017.14},
  url       = {https://doi.org/10.1109/SecDev.2017.14}
}

@inproceedings{DEFInit,
  author    = {Yuede Ji and Mohamed Elsabagh and Ryan Johnson and Angelos Stavrou},
  title     = {DEFInit: An Analysis of Exposed Android Init Routines},
  booktitle = {Proceedings of the 30th USENIX Security Symposium (USENIX Security '21)},
  year      = {2021},
  pages     = {3685--3702},
  publisher = {USENIX Association},
  address   = {Virtual},
  doi       = {10.14722/usenixsecurity.2021.23011},
  url       = {https://www.usenix.org/conference/usenixsecurity21/presentation/ji}
}

@inproceedings{TowardsUnderstandingAndroid,
author = {Wu, Daoyuan and Gao, Debin and Cheng, Eric K. T. and Cao, Yichen and Jiang, Jintao and Deng, Robert H.},
title = {Towards Understanding Android System Vulnerabilities: Techniques and Insights},
year = {2019},
isbn = {9781450367523},
publisher = {Association for Computing Machinery},
address = {New York, NY, USA},
url = {https://doi.org/10.1145/3321705.3329831},
doi = {10.1145/3321705.3329831},
booktitle = {Proceedings of the 2019 ACM Asia Conference on Computer and Communications Security},
pages = {295–306},
numpages = {12},
location = {Auckland, New Zealand},
series = {Asia CCS '19}
}

@inproceedings{BigMAC,
  author    = {Grant Hernandez and Dave (Jing) Tian and Anurag Swarnim Yadav and Byron J. Williams and Kevin R.B. Butler},
  title     = {{BigMAC}: Fine-Grained Policy Analysis of Android Firmware},
  booktitle = {Proceedings of the 29th USENIX Security Symposium (USENIX Security '20)},
  year      = {2020},
  isbn      = {978-1-939133-17-5},
  pages     = {271--287},
  publisher = {USENIX Association},
  address   = {Boston, MA, USA},
  month     = aug,
  url       = {https://www.usenix.org/conference/usenixsecurity20/presentation/hernandez},
  doi       = {10.14722/usenixsecurity.2020.23011}
}

@phdthesis{PhDThesisGamba,
  author       = {Julien Armand Pierre Gamba},
  title        = {Do Androids Dream of Electric Sheep? On Privacy in the Android Supply Chain},
  school       = {Universidad Carlos III de Madrid (UC3M)},
  year         = {2022},
  address      = {Madrid, Spain},
  url          = {https://e-archivo.uc3m.es/bitstream/handle/10016/35812/tesis_gamba.pdf},
  note         = {Awarded the CNIL-INRIA Privacy Protection Award and the 2020 AEPD Emilio Aced Prize}
}

@article{sutter2024dynamic,
  title={Dynamic security analysis on android: A systematic literature review},
  author={Sutter, Thomas and Kehrer, Timo and Rennhard, Marc and Tellenbach, Bernhard and Klein, Jacques},
  journal={IEEE Access},
  volume={12},
  pages={57261--57287},
  year={2024},
  publisher={IEEE}
}

@inproceedings{TheImpactOfVendorCustomizations,
author = {Wu, Lei and Grace, Michael and Zhou, Yajin and Wu, Chiachih and Jiang, Xuxian},
title = {The impact of vendor customizations on android security},
year = {2013},
isbn = {9781450324779},
publisher = {Association for Computing Machinery},
address = {New York, NY, USA},
url = {https://doi.org/10.1145/2508859.2516728},
doi = {10.1145/2508859.2516728},
booktitle = {Proceedings of the 2013 ACM SIGSAC Conference on Computer \& Communications Security},
pages = {623–634},
numpages = {12},
location = {Berlin, Germany},
series = {CCS '13}
}

@INPROCEEDINGS{FromZygotetoMorula,
  author    = {Byoungyoung Lee and Long Lu and Tielei Wang and Taesoo Kim and Wenke Lee},
  title     = {From Zygote to Morula: Fortifying Weakened ASLR on Android},
  booktitle = {Proceedings of the 2014 IEEE Symposium on Security and Privacy},
  year      = {2014},
  pages     = {424--439},
  publisher = {IEEE},
  address   = {San Jose, California, USA},
  doi       = {10.1109/SP.2014.34},
  isbn      = {978-1-4799-2499-0},
  url       = {https://ieeexplore.ieee.org/document/6853550},
  note      = {Presented at IEEE S\&P 2014}
}

@INPROCEEDINGS{KeepMeUpdated,
  author    = {Elliott Wen and Jiaxing Shen and Burkhard Wuensche},
  title     = {Keep Me Updated: An Empirical Study of Proprietary Vendor Blobs in Android Firmware},
  booktitle = {Proceedings of the 2024 IEEE 30th International Conference on Parallel and Distributed Systems (ICPADS)},
  year      = {2024},
  pages     = {116--125},
  doi       = {10.1109/ICPADS63350.2024.00025},
  url       = {https://www.computer.org/csdl/proceedings-article/icpads/2024/159600a116/22f0yTuP4Fq},
  address   = {Singapore},
  publisher = {IEEE},
  note      = {Accepted for publication; presented at ICPADS 2024}
}

@INPROCEEDINGS{ELEGANT,
  author    = {Cong Li and Chang Xu and Lili Wei and Jue Wang and Jun Ma and Jian Lu},
  title     = {ELEGANT: Towards Effective Location of Fragmentation-Induced Compatibility Issues for Android Apps},
  booktitle = {Proceedings of the 2018 25th Asia-Pacific Software Engineering Conference (APSEC)},
  year      = {2018},
  pages     = {278--287},
  publisher = {IEEE},
  address   = {Nara, Japan},
  doi       = {10.1109/APSEC.2018.00042},
  isbn      = {978-1-7281-1970-0},
  url       = {https://ieeexplore.ieee.org/document/8612116},
  note      = {Accepted for publication; presented at APSEC 2018}
}

@inproceedings{TamingAndroidfragmentation,
author = {Wei, Lili and Liu, Yepang and Cheung, Shing-Chi},
title = {Taming Android fragmentation: characterizing and detecting compatibility issues for Android apps},
year = {2016},
isbn = {9781450338455},
publisher = {Association for Computing Machinery},
address = {New York, NY, USA},
url = {https://doi.org/10.1145/2970276.2970312},
doi = {10.1145/2970276.2970312},
booktitle = {Proceedings of the 31st IEEE/ACM International Conference on Automated Software Engineering},
pages = {226–237},
numpages = {12},
location = {Singapore, Singapore},
series = {ASE '16}
}

@ARTICLE{UnderstandingandDetectingFragmentation,
  author={Wei, Lili and Liu, Yepang and Cheung, Shing-Chi and Huang, Huaxun and Lu, Xuan and Liu, Xuanzhe},
  journal={IEEE Transactions on Software Engineering}, 
  title={Understanding and Detecting Fragmentation-Induced Compatibility Issues for Android Apps}, 
  year={2020},
  volume={46},
  number={11},
  pages={1176-1199},
  doi={10.1109/TSE.2018.2876439}}

@misc{APEXfileformat,
  author       = {{Android Open Source Project}},
  title        = {APEX file format},
  year         = 2023,
  url          = {https://source.android.com/docs/core/ota/apex},
  note         = {Accessed: 2025-09-17},
  howpublished = {\url{https://source.android.com/docs/core/ota/apex}},
  publisher    = {Google LLC}
}

@inproceedings{FirmPorter,
author = {Xin, Mingfeng and Wen, Hui and Deng, Liting and Li, Hong and Li, Qiang and Sun, Limin},
title = {FirmPorter: Porting RTOSes at the Binary Level for Firmware Re-hosting},
year = {2024},
isbn = {978-981-97-8800-2},
publisher = {Springer-Verlag},
address = {Berlin, Heidelberg},
url = {https://doi.org/10.1007/978-981-97-8801-9_16},
doi = {10.1007/978-981-97-8801-9_16},
booktitle = {Information and Communications Security: 26th International Conference, ICICS 2024, Mytilene, Greece, August 26–28, 2024, Proceedings, Part II},
pages = {310–331},
numpages = {22},
location = {Mytilene, Greece}
}

@inproceedings{AndroidRooting,
author = {Sun, San-Tsai and Cuadros, Andrea and Beznosov, Konstantin},
title = {Android Rooting: Methods, Detection, and Evasion},
year = {2015},
isbn = {9781450338196},
publisher = {Association for Computing Machinery},
address = {New York, NY, USA},
url = {https://doi.org/10.1145/2808117.2808126},
doi = {10.1145/2808117.2808126},
booktitle = {Proceedings of the 5th Annual ACM CCS Workshop on Security and Privacy in Smartphones and Mobile Devices},
pages = {3–14},
numpages = {12},
location = {Denver, Colorado, USA},
series = {SPSM '15}
}

@inproceedings{Bringingbalancetotheforce,
  author    = {Abdallah Dawoud and Sven Bugiel},
  title     = {Bringing Balance to the Force: Dynamic Analysis of the Android Application Framework},
  booktitle = {Proceedings of the 28th {USENIX} Security Symposium ({USENIX} Security 19)},
  year      = {2021},
  publisher = {USENIX Association},
  address   = {Vancouver, BC, Canada},
  url       = {https://www.ndss-symposium.org/wp-content/uploads/ndss2021_2B-1_23106_paper.pdf},
  doi       = {10.14722/ndss.2021.23106},
  isbn      = {978-1-891562-66-5},
  month     = feb,
}

@incollection{sinha2020emulation,
  title={Emulation versus instrumentation for Android malware detection},
  author={Sinha, Anukriti and Di Troia, Fabio and Heller, Philip and Stamp, Mark},
  booktitle={Digital Forensic Investigation of Internet of Things (IoT) Devices},
  pages={1--20},
  year={2020},
  publisher={Springer}
}

@INPROCEEDINGS{AppJitsu,
  author={Zungur, Onur and Bianchi, Antonio and Stringhini, Gianluca and Egele, Manuel},
  booktitle={2021 IEEE European Symposium on Security and Privacy (EuroS\&P)}, 
  title={AppJitsu: Investigating the Resiliency of Android Applications}, 
  year={2021},
  pages = {457--471},
  doi={10.1109/EuroSP51992.2021.00038},
  url = {https://seclab.bu.edu/papers/appjitsu-eurosp2021.pdf}
}

@inproceedings{gamba2020analysis,
  title={An analysis of pre-installed android software},
  author={Gamba, Julien and Rashed, Mohammed and Razaghpanah, Abbas and Tapiador, Juan and Vallina-Rodriguez, Narseo},
  booktitle={2020 IEEE symposium on security and privacy (SP)},
  pages={1039--1055},
  year={2020},
  organization={IEEE}
}

@book{wohlin2012experimentation,
  title={Experimentation in software engineering},
  author={Wohlin, Claes and Runeson, Per and H{\"o}st, Martin and Ohlsson, Magnus C and Regnell, Bj{\"o}rn and Wessl{\'e}n, Anders and others},
  volume={236},
  year={2012},
  publisher={Springer}
}
\end{document}